\documentclass[12pt]{article}
\usepackage{pdproc,epsfig,url} 

\begin{document}

\renewcommand{\thefootnote}{\alph{footnote}}
  
\title{
RESULTS FROM HADROPRODUCTION EXPERIMENTS
}

\author{BORIS A. POPOV}

\address{ LPNHE, 4 place Jussieu, 75005, Paris \&
DLNP, JINR, Joliot-Curie 6, 141980, Dubna\\
 {\rm E-mail: Boris.Popov@cern.ch}}

\abstract{
The hadroproduction experiments HARP, MIPP and NA61 (SHINE) as well as their
implications for neutrino physics are discussed. 
HARP measurements have already been used for precise predictions of
neutrino beams in K2K and MiniBooNE/SciBooNE experiments and are also
being used to improve the atmospheric neutrino flux predictions and to
help in the optimization of neutrino factory and super-beam
designs. 
The MIPP experiment has nearly finalized measurements of hadron production
from the NuMI target used in the MINOS experiment.  
First measurements released recently by the NA61 (SHINE) experiment 
are of significant
importance for a precise prediction of the J-PARC neutrino beam
used for the first stage of the T2K experiment. 
All three experiments -- HARP, MIPP and NA61 -- 
provide also a large amount of input 
for validation and tuning of
hadron production models in Monte-Carlo generators.
}
   
\normalsize\baselineskip=15pt

\section{The HARP experiment}

The HARP experiment~\cite{ref:harpTech} at the CERN PS
was designed to make measurements of hadron yields from a large range
of nuclear targets from hydrogen to lead and for incident particle momenta 
from 1.5~GeV/c to 15~GeV/c.
The main motivations are the measurement of pion yields for a quantitative
design of the proton driver of a future neutrino
factory~\cite{ref:nufact}, 
a substantial improvement in the calculation of 
the atmospheric neutrino flux~\cite{ref:atm_nu_flux}
and the measurement of particle yields as input for the flux
calculation in accelerator neutrino experiments,
such as K2K~\cite{ref:k2k,ref:k2kfinal},
MiniBooNE~\cite{ref:miniboone} and SciBooNE~\cite{ref:sciboone}.
In addition to these specific aims, the data provided by HARP are
valuable for validation and tuning of hadron production models 
used in simulation programs. 
The HARP experiment is described in more details below in order to illustrate 
some common features of modern hadron production experiments. 

To provide a large angular and momentum coverage of the produced charged
particles the HARP experiment
 makes use of a large-acceptance spectrometer consisting of a
 forward and large-angle detection system.
 A detailed
 description of the experimental apparatus can be found in Ref.~\cite{ref:harpTech}.
 The forward spectrometer --- 
 based on large area drift chambers~\cite{ref:NOMAD_NIM_DC} and a dipole magnet
 complemented by a set of detectors for particle identification (PID): 
 a time-of-flight wall~\cite{barichello} (TOFW), a large Cherenkov detector (CHE) 
 and an electromagnetic calorimeter  ---
 covers polar angles up to 250~mrad which
 is well matched to the angular range of interest for the
 measurement of hadron production to calculate the properties of
 conventional neutrino beams.
 The large-angle spectrometer --- based on a Time Projection Chamber (TPC) 
 located inside a solenoidal magnet ---
 has a large acceptance in the momentum
 and angular range for the pions relevant to the production of the
 muons in a neutrino factory.
 It covers a large majority of the pions accepted in the focusing
 system of a typical design.
 The neutrino beam of a neutrino factory originates from
 the decay of muons which are in turn the decay products of pions.

A full set of data collected by the HARP experiment with thin 
(5\% of nuclear interaction length, $\lambda_{\mathrm{I}}$) 
and cryogenic targets have 
been analyzed and published~\cite{ref:alPaper,ref:bePaper,ref:harp:carbonfw,ref:harp:o2n2,ref:harp:forward_pi,ref:harp:tantalum,ref:harp:cacotin,ref:harp:bealpb,ref:harp:la,ref:harp:la:pions,ref:harp:forward_protons,ref:harp:forward_protons_in__protons_and_pions}.
Those results cover all the physics subjects discussed above.

\subsection{Results obtained with the HARP forward spectrometer}

A detailed description of established experimental techniques 
for the data analysis in the HARP forward spectrometer can be found 
in Refs.~\cite{ref:alPaper,ref:pidPaper,ref:bePaper,ref:harp:carbonfw}.
Only a brief summary is given here.

The absolutely normalized double-differential cross-section for a process 
like $p + \mbox{Target} \rightarrow \pi^+ + X$
can be expressed in bins of pion kinematic variables in the
laboratory frame (momentum, $p_{\pi}$, and polar angle, $\theta_{\pi}$), as 
\begin{equation}
  \frac{d^2\sigma^{\pi^{+}}}{dpd\Omega}(p_{\pi},\theta_{\pi}) = \frac{A}{N_{\mbox{A}}\rho t}\cdot\frac{1}{\Delta p\Delta\Omega}\cdot\frac{1}{N_{\mbox{pot}}}
    \cdot N^{\pi^{+}}(p_{\pi},\theta_{\pi}) \ ,
    \label{eq:truexsec}
\end{equation}
where
$\frac{d^2\sigma^{\pi^{+}}}{dpd\Omega}$ is the cross-section in 
\ensuremath{\mbox{cm}^2/(\mbox{GeV/c})/\mbox{srad}} 
for each ($p_{\pi},\theta_{\pi}$) bin covered in the analysis;
$\frac{A}{N_{\mbox{\tiny A}}\rho}$ is the reciprocal of the number density of
  target nuclei; 
$t$ is the thickness of the 
target along the beam direction; 
$\Delta p$ and $\Delta \Omega$ are the bin sizes in momentum and
solid angle ($\Delta p = p_{max}-p_{min};
\ \Delta \Omega = 2\pi(cos(\theta_{min}) - cos(\theta_{max}))$); 
$N_{\mbox{pot}}$ is the number of protons on target after event
  selection cuts;
$N^{\pi^{+}}(p_{\pi},\theta_{\pi})$ is the yield of positive pions in bins
of true momentum and angle in the laboratory frame.
Eq.~(\ref{eq:truexsec}) can be generalized to give the inclusive
cross-section for a particle of type $\alpha$      
\begin{equation}
  \frac{d^2\sigma^{\alpha}}{dpd\Omega}(p,\theta) = \frac{A}{N_{\mbox{A}}\rho t}\cdot\frac{1}{\Delta p\Delta\Omega}\cdot\frac{1}{N_{\mbox{pot}}}
  \cdot M^{-1}_{p\theta\alpha p^{'}\theta^{'}\alpha^{'}}\cdot N^{\alpha^{'}}(p^{'},\theta^{'}) \ ,
  \label{eq:recxsec}
\end{equation}
where reconstructed quantities are marked with a prime and 
$M^{-1}_{p\theta\alpha p^{'}\theta^{'}\alpha^{'}}$ is the
inverse of a matrix which fully describes the migrations between bins
of true and 
reconstructed quantities, namely: lab frame momentum, p, lab frame
angle, $\theta$, and particle type, $\alpha$.

There is a background associated with beam protons interacting in materials 
other than the nuclear target (parts of the detector, air, etc.).  
These events 
are
subtracted by using data collected without the nuclear target 
in place after normalization to the same 
number of protons on target. This procedure is referred to as 
the `empty target subtraction'.  

The event selection is performed in the following way:
a good event is required to have a single, well reconstructed and
identified beam particle 
impinging on the nuclear target.  
A downstream trigger in the forward trigger plane (FTP) 
is also required to record the event, necessitating an additional set of unbiased,
pre-scaled triggers for absolute normalization of the  
cross-section.  These pre-scaled triggers (e.g. 1/64) 
are subject to exactly the same selection 
criteria for a `good' beam particle as the event triggers allowing the
efficiencies of the selection to cancel, thus adding no additional
systematic uncertainty to the absolute normalization of the result. 
Secondary track selection criteria have been optimized to ensure the
quality of the  momentum reconstruction as well as a clean time-of-flight 
measurement while maintaining high reconstruction and particle identification 
efficiencies~\cite{ref:pidPaper,ref:bePaper}.

The first HARP physics publication~\cite{ref:alPaper} 
reported measurements of the
$\pi^+$ production cross-section from a thin 
5\%~$\lambda_{\mathrm{I}}$ aluminum target 
at 12.9~GeV/c proton momentum. 
This corresponds to the energies of the KEK PS
and the target material used by the K2K experiment.  
The results obtained in Ref.~\cite{ref:alPaper} were
subsequently applied to the final neutrino oscillation analysis 
of K2K~\cite{ref:k2kfinal}, allowing a significant reduction 
of the dominant systematic error associated with the calculation of
the so-called far-to-near ratio from 5.1\% to 2.9\%
(see~\cite{ref:alPaper} and~\cite{ref:k2kfinal} for a detailed discussion) 
and thus an increased K2K sensitivity to the oscillation signal. 

The next HARP goal was to contribute to the understanding of 
the MiniBooNE and SciBooNE neutrino fluxes. 
They are both produced by the Booster Neutrino Beam at 
Fermilab which originates from protons accelerated to 8.9~GeV/c by 
the booster before being collided against a beryllium target. 
A fundamental input for the calculation 
of the resulting $\nu_\mu$ flux is the measurement of the $\pi^+$ cross-sections 
from a thin 5\%~$\lambda_{\mathrm{I}}$
beryllium target at exactly 8.9~GeV/c proton momentum.

The double-differential cross-section for the
production of positively charged pions from collisions of 8.9~GeV/c protons 
with beryllium have been measured in the kinematic range from 
$0.75 \ \mbox{GeV/c} \leq p_{\pi} \leq 6.5$~GeV/c and 
$0.030 \ \mbox{rad} \leq \theta_{\pi} \leq 0.210$~rad,
subdivided into 13 momentum and 6 angular bins~\cite{ref:bePaper}.  
A typical total uncertainty of 9.8\% on the double-differential cross-section values 
and a 4.9\% uncertainty on the total integrated cross-section are obtained.    
These HARP results have been used for neutrino flux 
predictions~\cite{miniboone_flux} in the 
MiniBooNE~\cite{miniboone_osc} and SciBooNE
experiments~\cite{ref:sciboone}.

Sanford and Wang~\cite{ref:SW} have developed an empirical parametrization for
describing the production cross-sections of mesons in proton-nucleus 
interactions (e.g. $\hbox{p+A}\to \pi^++X$).
This SW parametrization has the functional form: 
\begin{equation}
\label{eq:swformula}
\frac{d^2\sigma}{dpd\Omega}(p,\theta) =
 \exp [ c_{1}-c_{3}\frac{p^{c_{4}}}{p_{\hbox{\footnotesize beam}}^{c_{5}}}-c_{6}
 \theta (p-c_{7} p_{\hbox{\footnotesize {beam}}} \cos^{c_{8}}\theta ) ] p^{c_{2}}
 (1-\frac{p}{p_{\hbox{\footnotesize beam}}}) \ ,
\end{equation} 
where $X$ denotes any system of other particles in the final state;
$p_{\hbox{\footnotesize {beam}}}$ is
the proton beam momentum in GeV/c; $p$ and $\theta$ are the $\pi^+$
momentum and angle in units of GeV/c and radians, respectively; 
$d^2\sigma/(dpd\Omega)$ is expressed in units of mb/(GeV/c\ sr);
$d\Omega\equiv 2\pi\ d(\cos\theta )$; 
and the parameters $c_1,\ldots ,c_8$
are obtained from fits to meson production data. 
HARP data have been fitted using this empirical SW parametrization.
This is a useful tool to compare and combine different data sets.
In~\cite{ref:harp:forward_protons}
a global SW parametrization is provided 
for forward production of charged pions
as an approximation of all the studied datasets. 
It can serve as a tool for quick yields estimates.

\begin{figure}
\epsfig{file=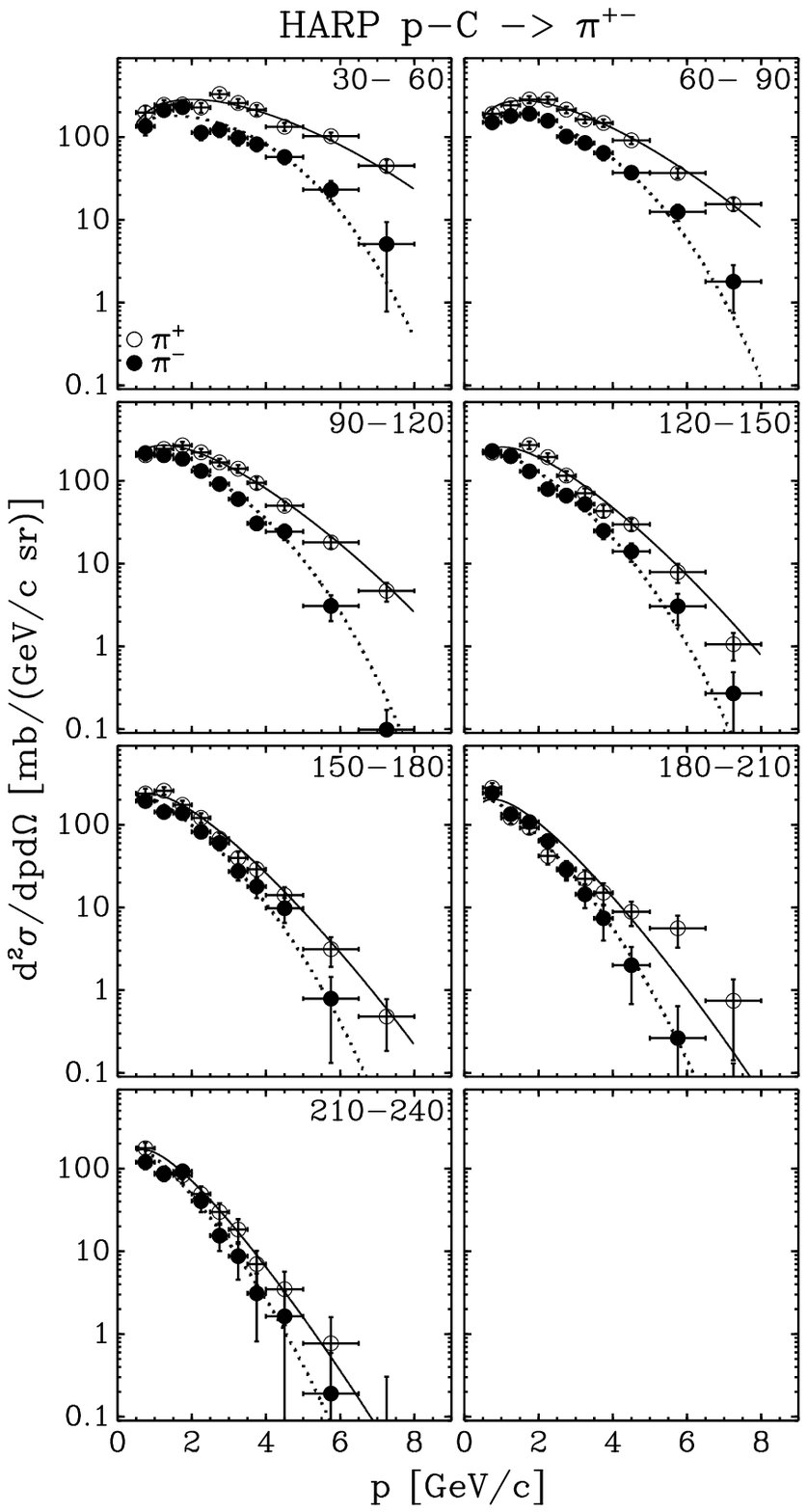, height=7.5cm, width=0.325\textwidth}
\epsfig{file=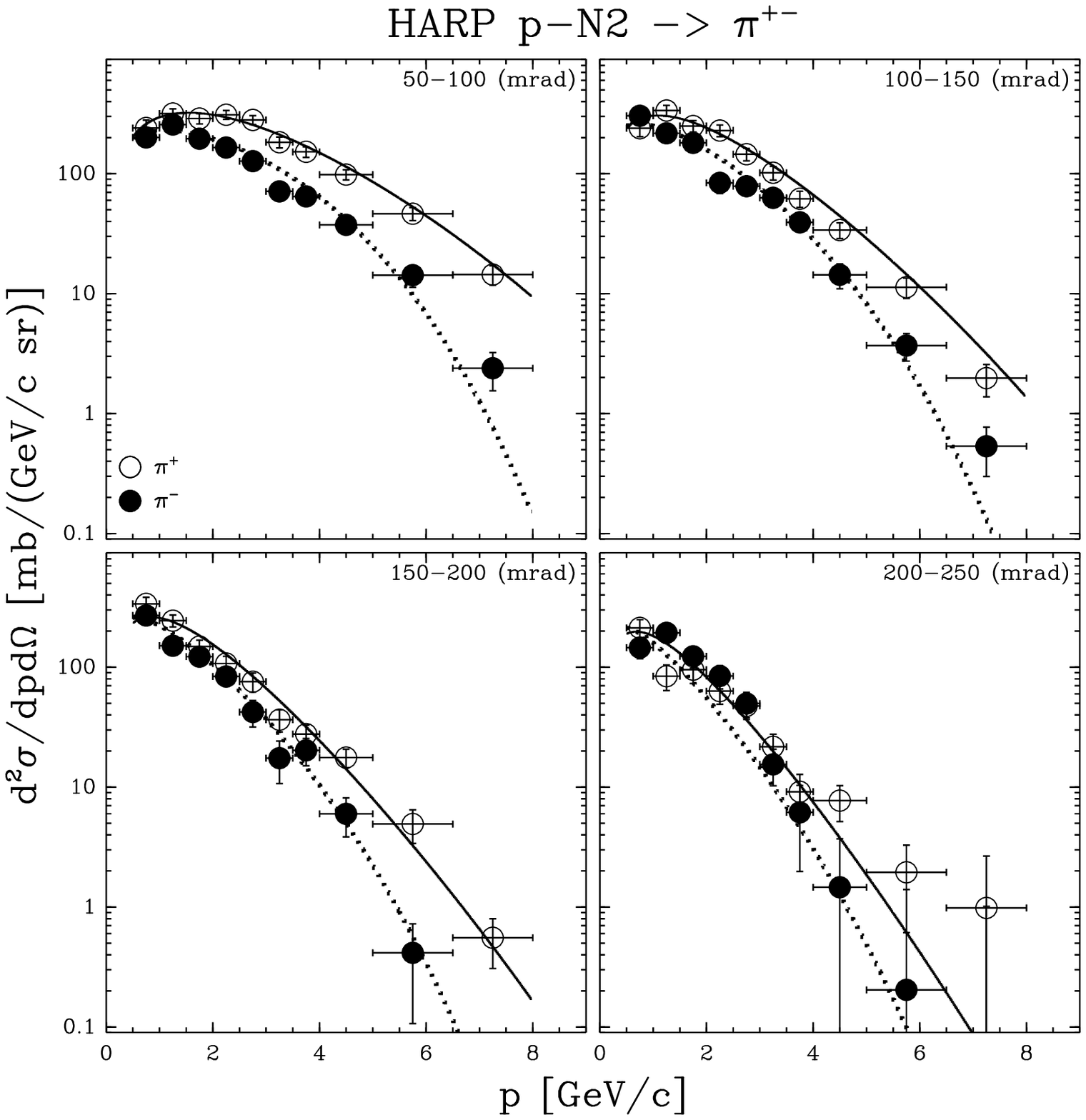, height=7.5cm, width=0.325\textwidth}
\epsfig{file=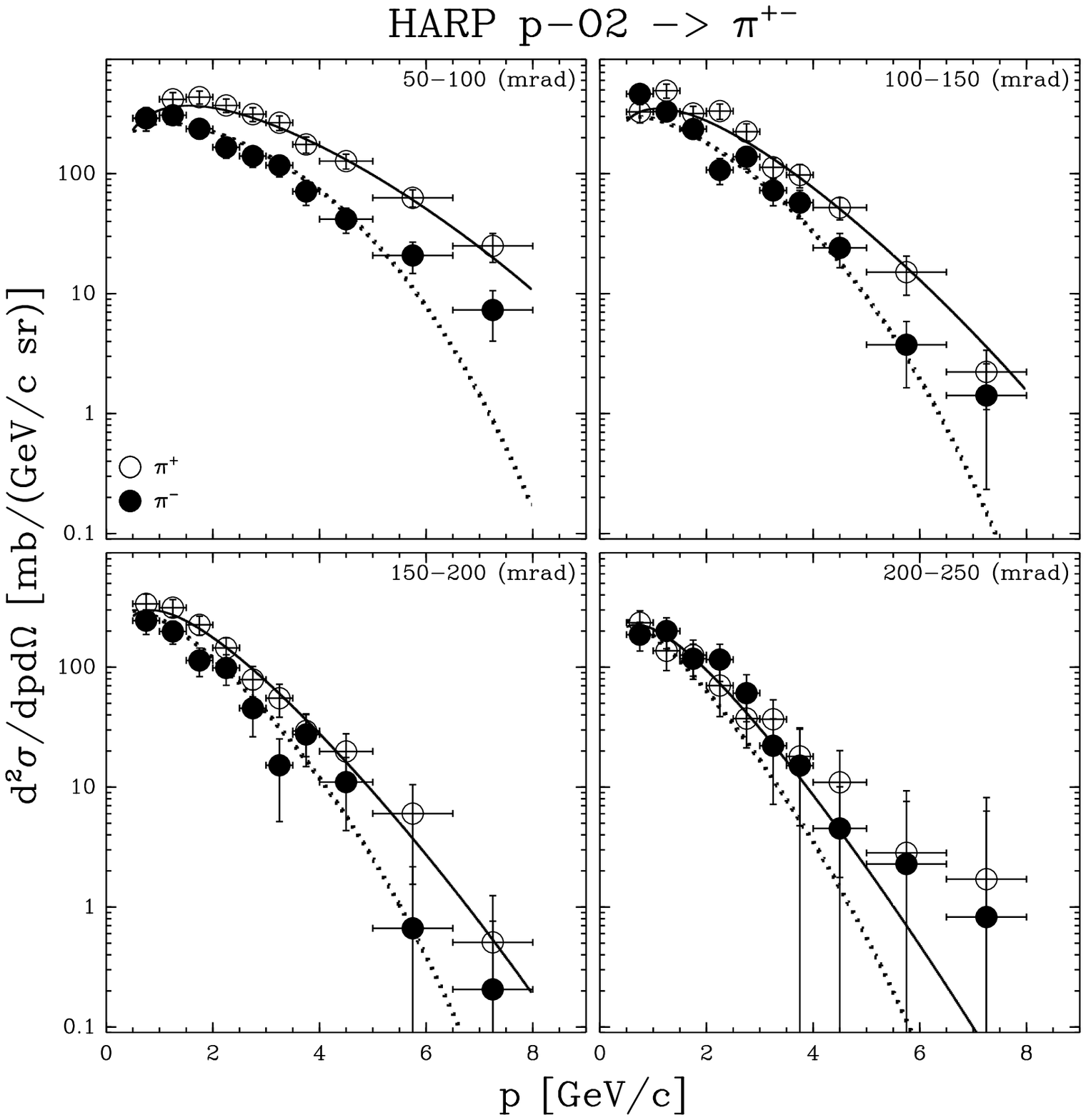, height=7.5cm, width=0.325\textwidth}
\caption{
  \label{fig:fw_carbon}
  Measurements of the double-differential production
  cross-sections of 
  $\pi^+$ (open circles) and $\pi^-$ (closed circles)
  from 12~GeV/c protons on carbon (left), N$_2$ (middle) and 
  O$_2$ (right) target as a function 
  of pion momentum, p, in bins of pion angle, $\theta$, in the 
  laboratory frame.  
  Different panels show different angular bins.
  The error bars shown include statistical
  errors and all (diagonal) systematic errors. 
  The curves show the SW parametrization 
  of Eq.~(\ref{eq:swformula}) with parameters given 
  in Ref.~$^{13)}$ 
  (solid line for $\pi^+$ and dashed line for $\pi^-$). 
}  
\end{figure}

The next HARP analysis 
was devoted to
the measurement 
of the double-differential production cross-section of $\pi^\pm$ 
in the collision of 12~GeV/c protons (see Fig.~\ref{fig:fw_carbon}) and pions 
with a thin 5\%~$\lambda_{\mathrm{I}}$ 
carbon target~\cite{ref:harp:carbonfw}.
These measurements are important for a precise calculation 
of the atmospheric neutrino
flux and for a prediction of the development of extended air showers.
Simulations predict that collisions of protons with a carbon target
are very similar to proton interactions with air. This hypothesis
could be directly tested with the HARP data. Measurements with
cryogenic O$_2$ and N$_2$ targets~\cite{ref:harp:o2n2}, 
also shown in Fig.~\ref{fig:fw_carbon},
confirm that p-C data can indeed be used to
predict pion production in p-O$_2$ and p-N$_2$ interactions.

\begin{figure}
\epsfig{file=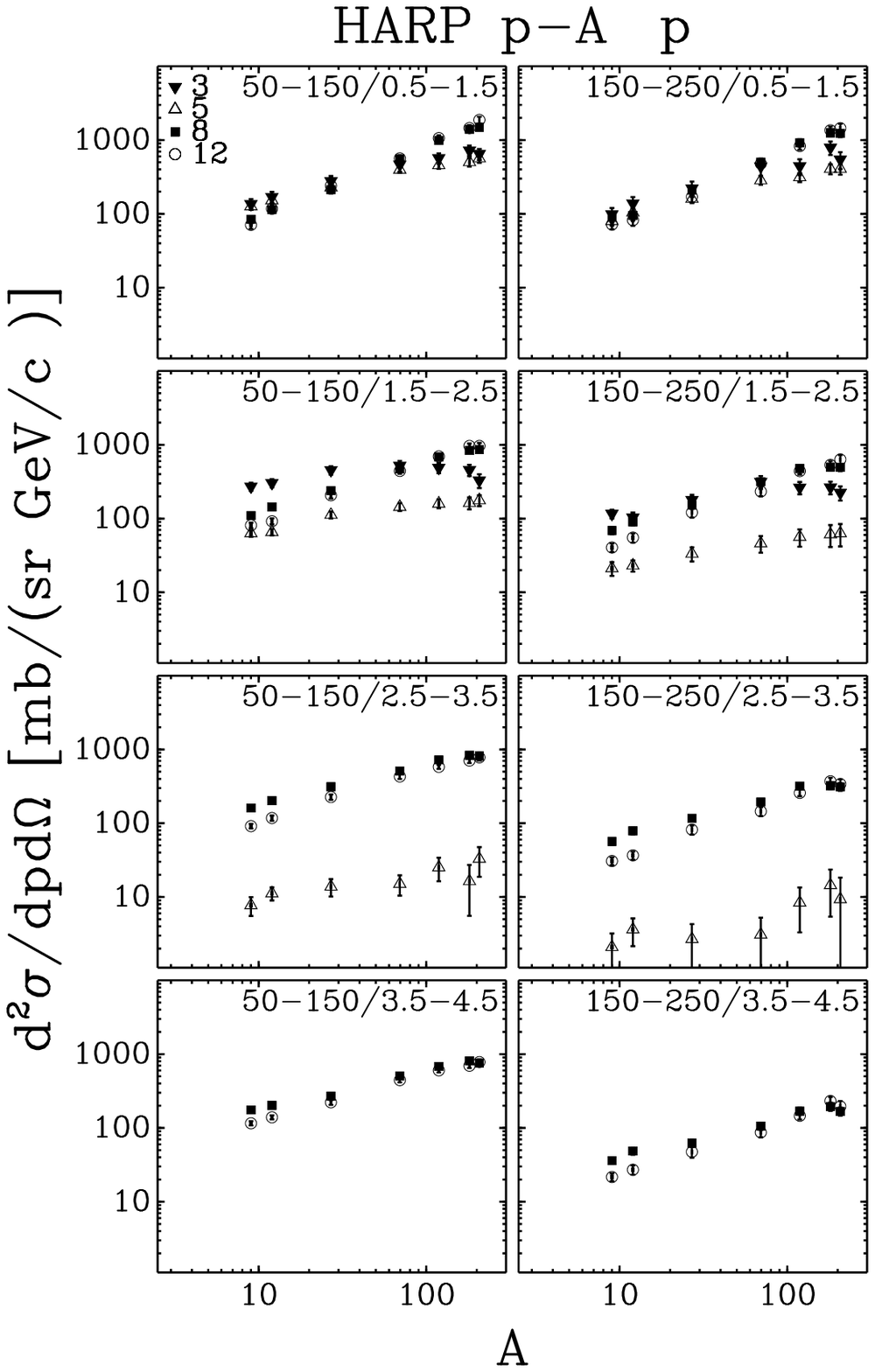, width=0.32\textwidth}
\epsfig{file=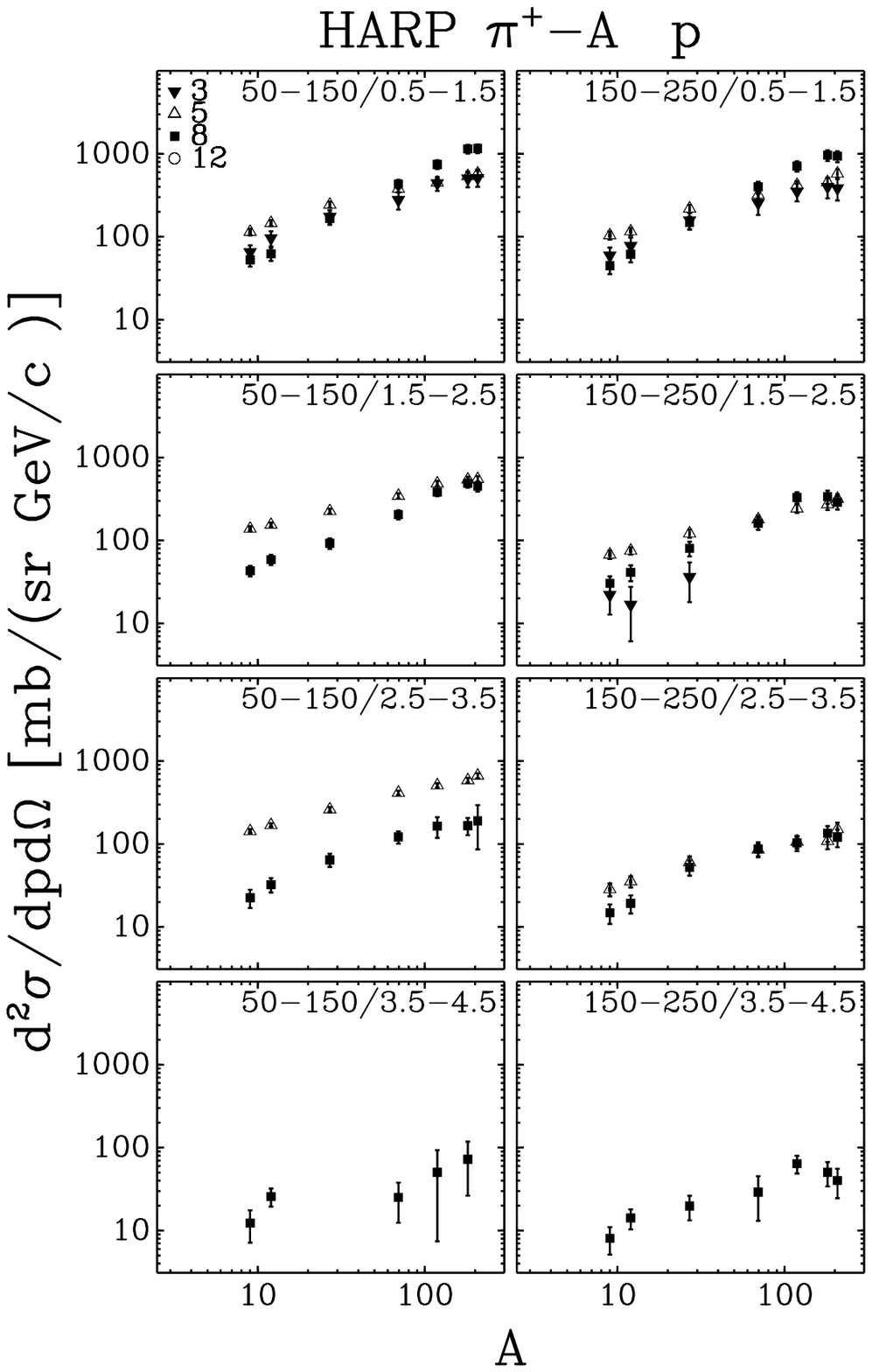, width=0.32\textwidth}
\epsfig{file=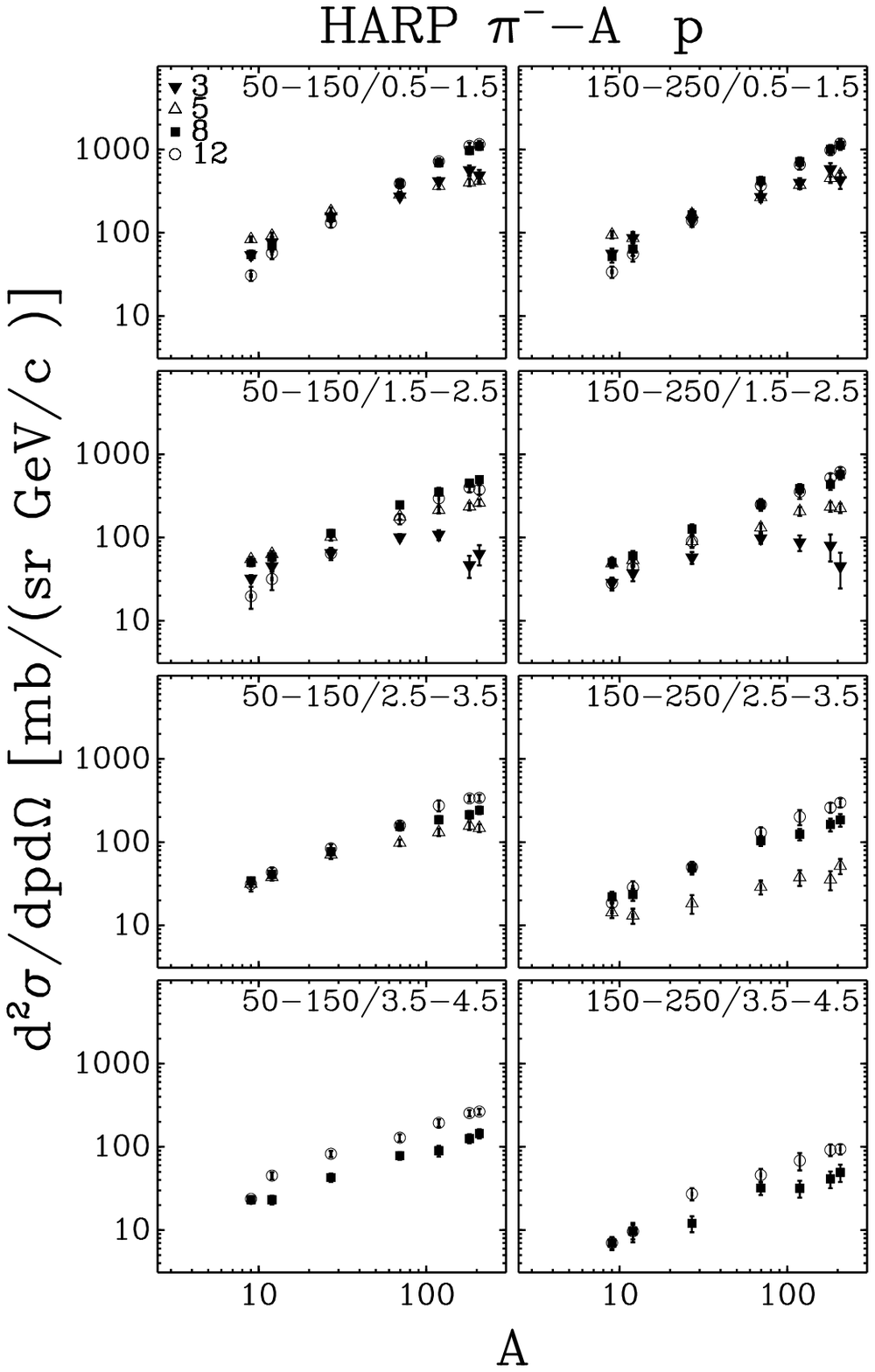, width=0.32\textwidth}
\caption{
\label{fig:proton}
 The dependence on the atomic number $A$ 
 of the forward proton production yields in p--$A$ and $\pi^\pm$--$A$
 ($A$ = Be, C, Al, Cu, Sn, Ta, Pb) 
 interactions averaged over two forward angular regions
 ($0.05~\mbox{rad} \leq \theta < 0.15~\mbox{rad}$ and
  $0.15~\mbox{rad} \leq \theta < 0.25~\mbox{rad}$)
 and four momentum regions
  ($0.5~\mbox{GeV}/c \leq p < 1.5~\mbox{GeV}/c$,
   $1.5~\mbox{GeV}/c \leq p < 2.5~\mbox{GeV}/c$,
   $2.5~\mbox{GeV}/c \leq p < 3.5~\mbox{GeV}/c$ and
   $3.5~\mbox{GeV}/c \leq p < 4.5~\mbox{GeV}/c$), for the four different
  incoming beam momenta (3, 5, 8 and 12~GeV/c).
}
\end{figure}

It is important to emphasize that a systematic campaign of
measurements has been performed by HARP: all thin target data taken with
pion beams are analyzed and published now~\cite{ref:harp:forward_pi}.
Similar results have also been obtained for incoming 
protons~\cite{ref:harp:forward_protons}.
Moreover, results on forward proton production 
in p-$A$ and $\pi^\pm$-$A$ interactions have been published 
recently~\cite{ref:harp:forward_protons_in__protons_and_pions}.
Fig.~\ref{fig:proton} shows the dependence on the atomic number $A$ of the
forward proton yields in p--$A$ and $\pi^\pm$--$A$ interactions 
averaged over two forward angular regions
 ($0.05~\mbox{rad} \leq \theta < 0.15~\mbox{rad}$ and
  $0.15~\mbox{rad} \leq \theta < 0.25~\mbox{rad}$)
 and four momentum regions
  ($0.5~\mbox{GeV}/c \leq p < 1.5~\mbox{GeV}/c$,
   $1.5~\mbox{GeV}/c \leq p < 2.5~\mbox{GeV}/c$,
   $2.5~\mbox{GeV}/c \leq p < 3.5~\mbox{GeV}/c$ and
   $3.5~\mbox{GeV}/c \leq p < 4.5~\mbox{GeV}/c$), for the four different
  incoming beam momenta (3, 5, 8 and 12~GeV/c).

The advantage of all these measurements is that they are performed with
the same detector, thus related systematic uncertainties are minimized.

\subsection{Results obtained with the HARP large-angle spectrometer}

\begin{figure}
\epsfig{file=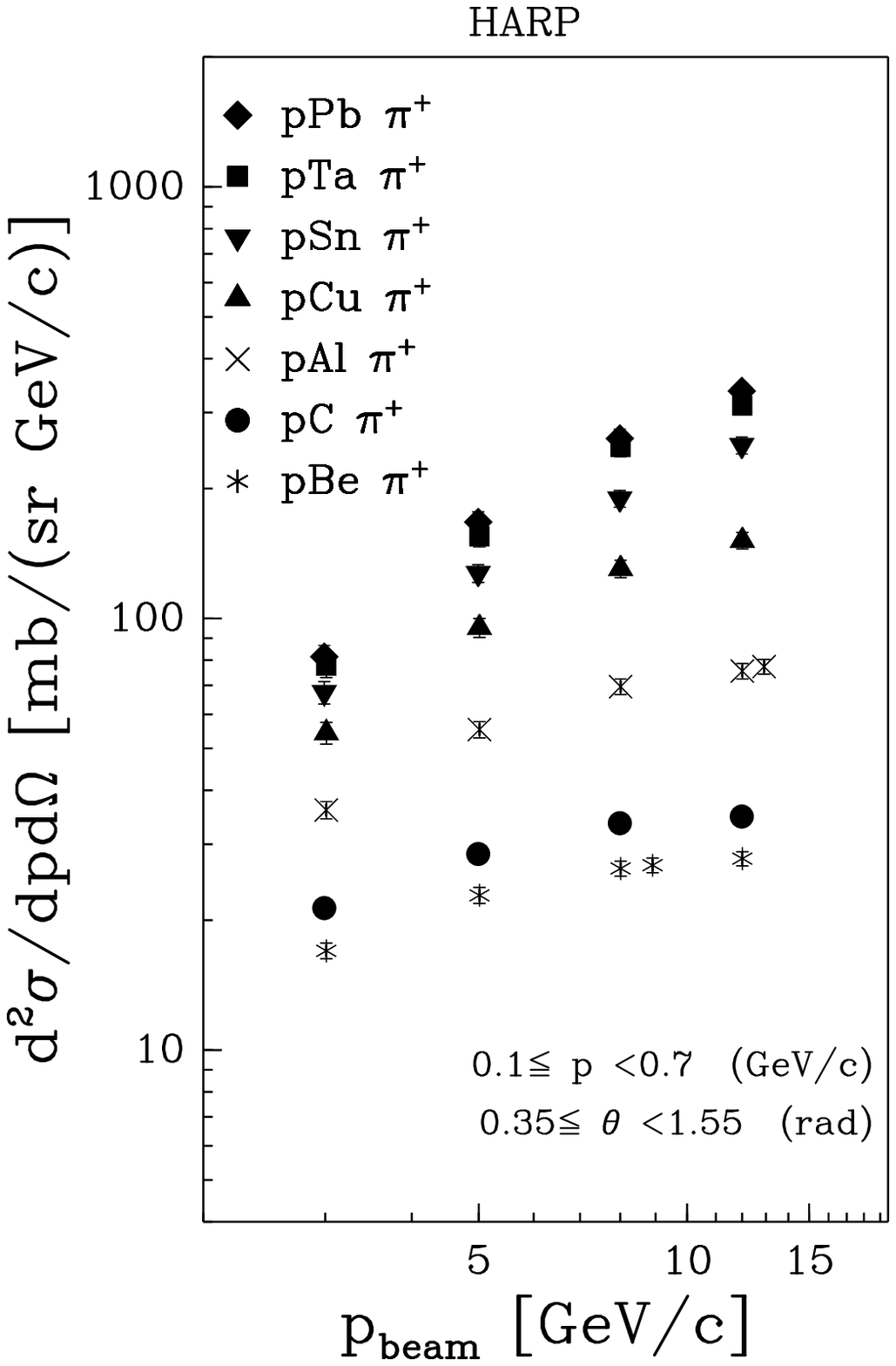, height=7.cm, width=0.45\textwidth}
\epsfig{file=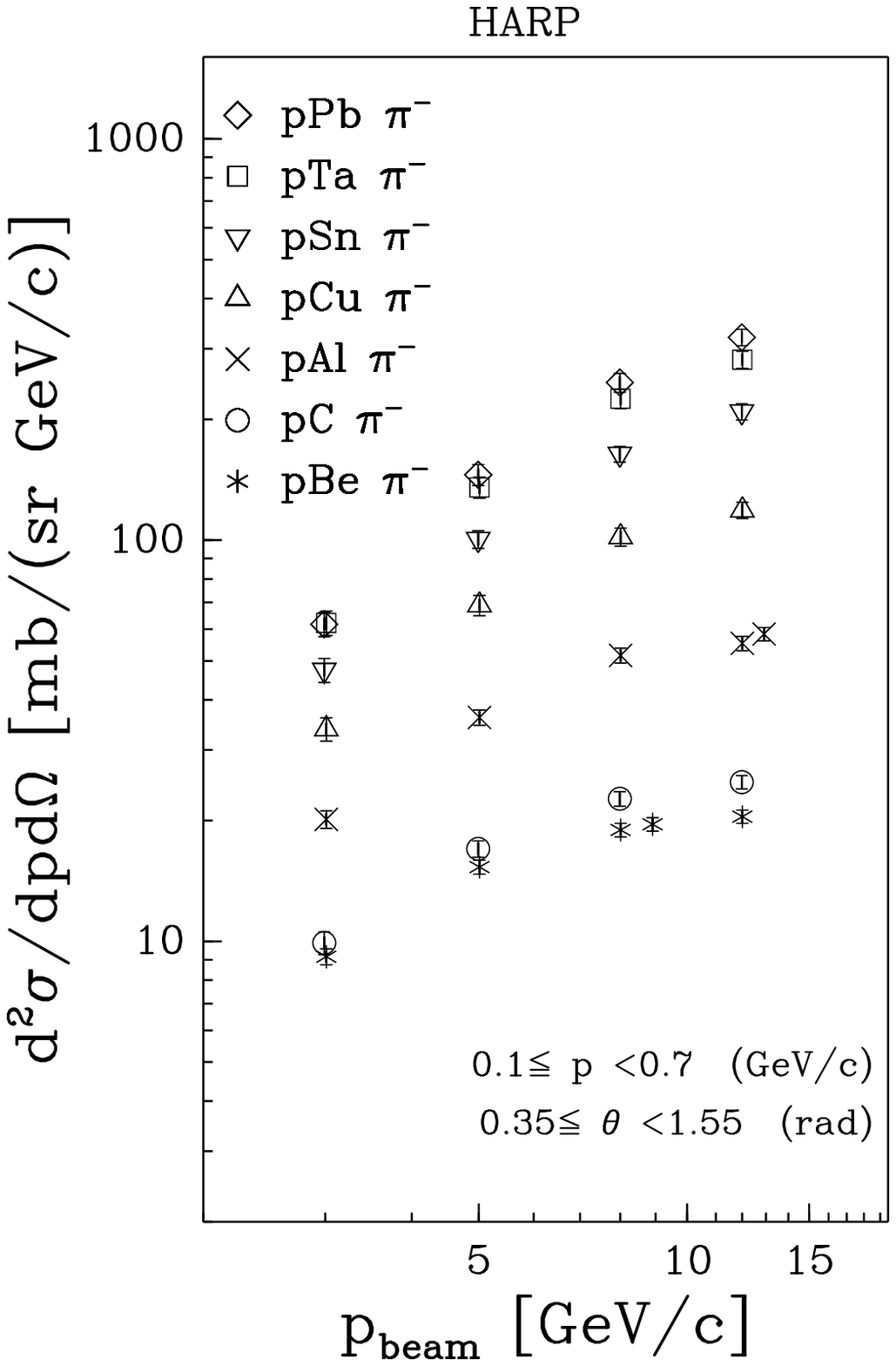, height=7.cm, width=0.45\textwidth}
\caption{
\label{fig:xs-trend}
 The dependence on the beam momentum of the $\pi^+$ (left) and $\pi^-$ (right)
  production yields in p--$A$ ($A$ = Be, C, Al, Cu, Sn, Ta, Pb)
 interactions averaged over the forward angular region
 ($0.350~\mbox{rad} \leq \theta < 1.550~\mbox{rad}$)
 and momentum region $100~\mbox{MeV}/c \leq p < 700~\mbox{MeV}/c$.
 The results are given in arbitrary units, with a consistent scale
 between the left and right panel.
 Data points for different target nuclei and equal momenta are slightly
 shifted horizontally with respect to each other to increase the visibility.
}
\end{figure}

  First results on the measurements of 
  the double-differential cross-section for the production
  of charged pions in p-$A$ collisions emitted at large
  angles from the incoming beam direction have been obtained 
  for the full set of available nuclear targets
  using an initial part of the accelerator spill to avoid distortions 
  in the TPC~\cite{ref:harp:tantalum,ref:harp:cacotin,ref:harp:bealpb}. 
  The pions were produced by proton beams in a momentum range from
  3~GeV/c to  12~GeV/c hitting a target with a thickness of
  5\%~$\lambda_{\mathrm{I}}$.  
  The angular and momentum range covered by the experiment 
  ($100~\mbox{MeV/c} \le p < 800~\mbox{MeV/c}$ and 
  $0.35~\mbox{rad} \le \theta <2.15~\mbox{rad}$)
  is of particular importance for the design of a neutrino factory.
  Track recognition, momentum determination and particle
  identification were all performed based on the measurements made with
  the TPC (see~\cite{ref:harp:tantalum} for more details). 

After the development of a special algorithm to correct for
the TPC dynamic distortions~\cite{ref:dyndist} and validation of the
TPC performance with benchmarks based on real data~\cite{ref:tpcmom},
full spill data have been analyzed and published for
incoming protons~\cite{ref:harp:la} and pions~\cite{ref:harp:la:pions}. 
Fig.~\ref{fig:xs-trend} illustrates the dependences of the measured
charged pion yields on the beam momentum 
in p--$A$ ($A$ = Be, C, Al, Cu, Sn, Ta, Pb)
interactions averaged over the forward angular region
($0.350~\mbox{rad} \leq \theta < 1.550~\mbox{rad}$)
and momentum region $100~\mbox{MeV}/c \leq p < 700~\mbox{MeV}/c$.

An additional analysis has also been performed on 
the comparison of the pion yields measured with the
short (5\%~$\lambda_{\mathrm{I}}$) and
long (100\%~$\lambda_{\mathrm{I}}$) 
nuclear targets~\cite{ref:harp:long_targets}.  

All HARP measurements described above have been compared 
with predictions of Monte-Carlo models
available within GEANT4~\cite{ref:geant4} and MARS~\cite{ref:mars}, as well as
with the GiBUU model~\cite{ref:GiBUU}. Some models provide a
reasonable description of HARP data in some regions~\cite{ref:harp:carbonfw,ref:harp:o2n2,ref:harp:forward_pi,ref:harp:tantalum,ref:harp:cacotin,ref:harp:bealpb,ref:harp:la,ref:harp:la:pions,ref:harp:forward_protons,ref:harp:forward_protons_in__protons_and_pions}, while there is no
model which would describe all aspects of the data. 

Note that tables with HARP results in text format are available 
from the DURHAM database~\cite{DURHAM}.

\section{The MIPP experiment}

The MIPP experiment at Fermilab took data in 2005. Some of these data are
now analysed and published~\cite{MIPP_neutrons,MIPP_kaon}. 
The analysis of hadron yields in 120~GeV/c proton
interactions with the NuMI carbon target is nearly complete. These results
could be used to improve precision of neutrino flux predictions in the MINOS
experiment~\cite{MINOS}.

There is also a recent proposal for an upgrade of 
the MIPP experiment~\cite{MIPP_upgrade}.

\section{The NA61 (SHINE) experiment}

The physics program of the NA61 or SHINE 
(where SHINE $\equiv$ SPS Heavy Ion and Neutrino Experiment)
experiment at the CERN SPS~\cite{proposal} consists
of three main subjects. In the first stage of data taking 
measurements of hadron production in hadron-nucleus interactions
needed for neutrino (T2K~\cite{t2k}) 
and cosmic-ray (Pierre Auger~\cite{auger} and KASCADE~\cite{kaskade})
experiments are performed. 
At later stages of the NA61 experiment
hadron production in proton-proton and proton-nucleus interactions
needed as reference data for a better understanding of nucleus-nucleus
reactions will be studied, 
energy dependence of hadron
production properties 
will be measured in p-p and p-Pb interactions as well as in
nucleus-nucleus collisions, with the aim
to identify the properties of the onset of deconfinement and find
evidence for the critical point of strongly interacting matter.

\begin{figure}[htb]
\epsfig{file=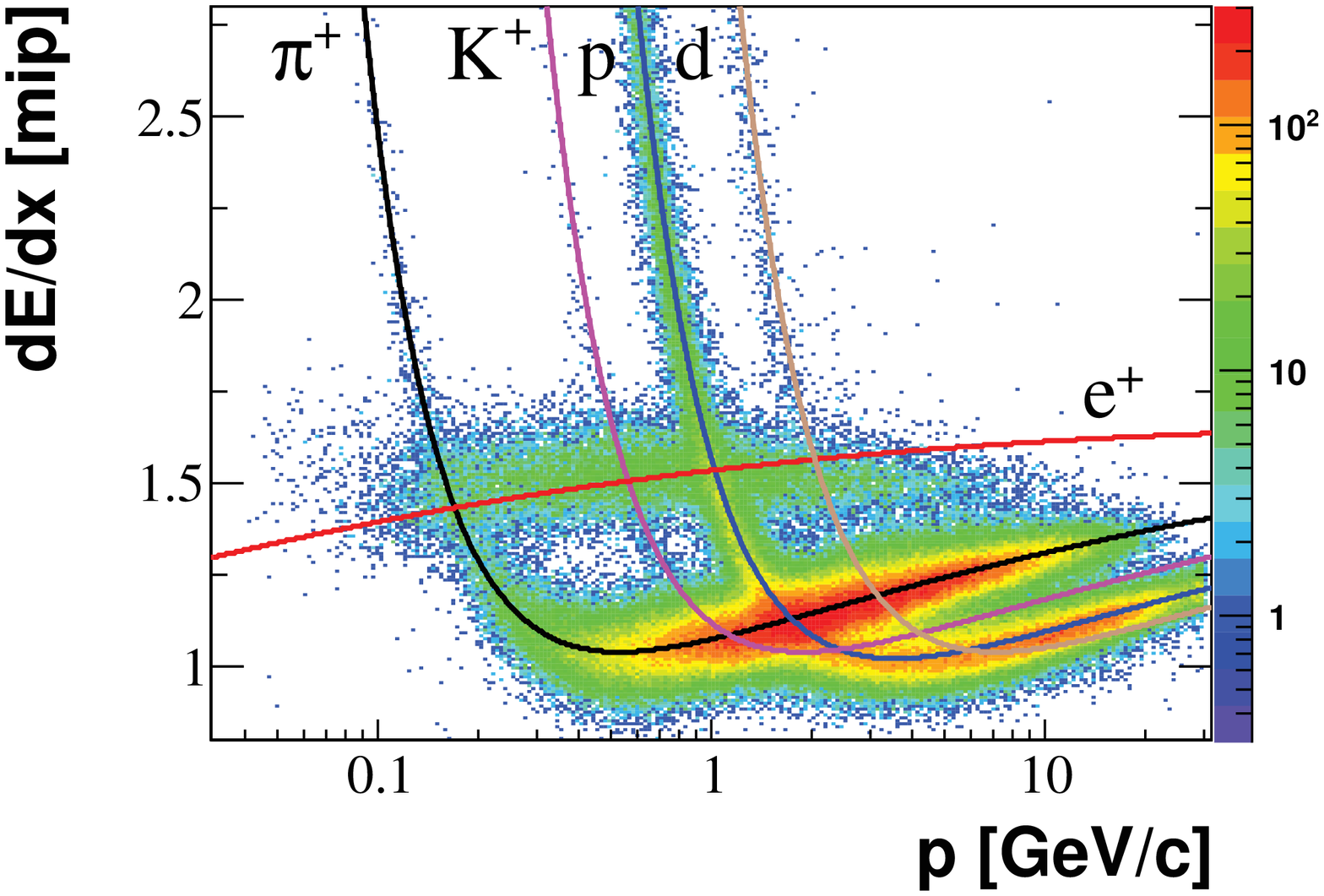,width=0.325\textwidth}
\epsfig{file=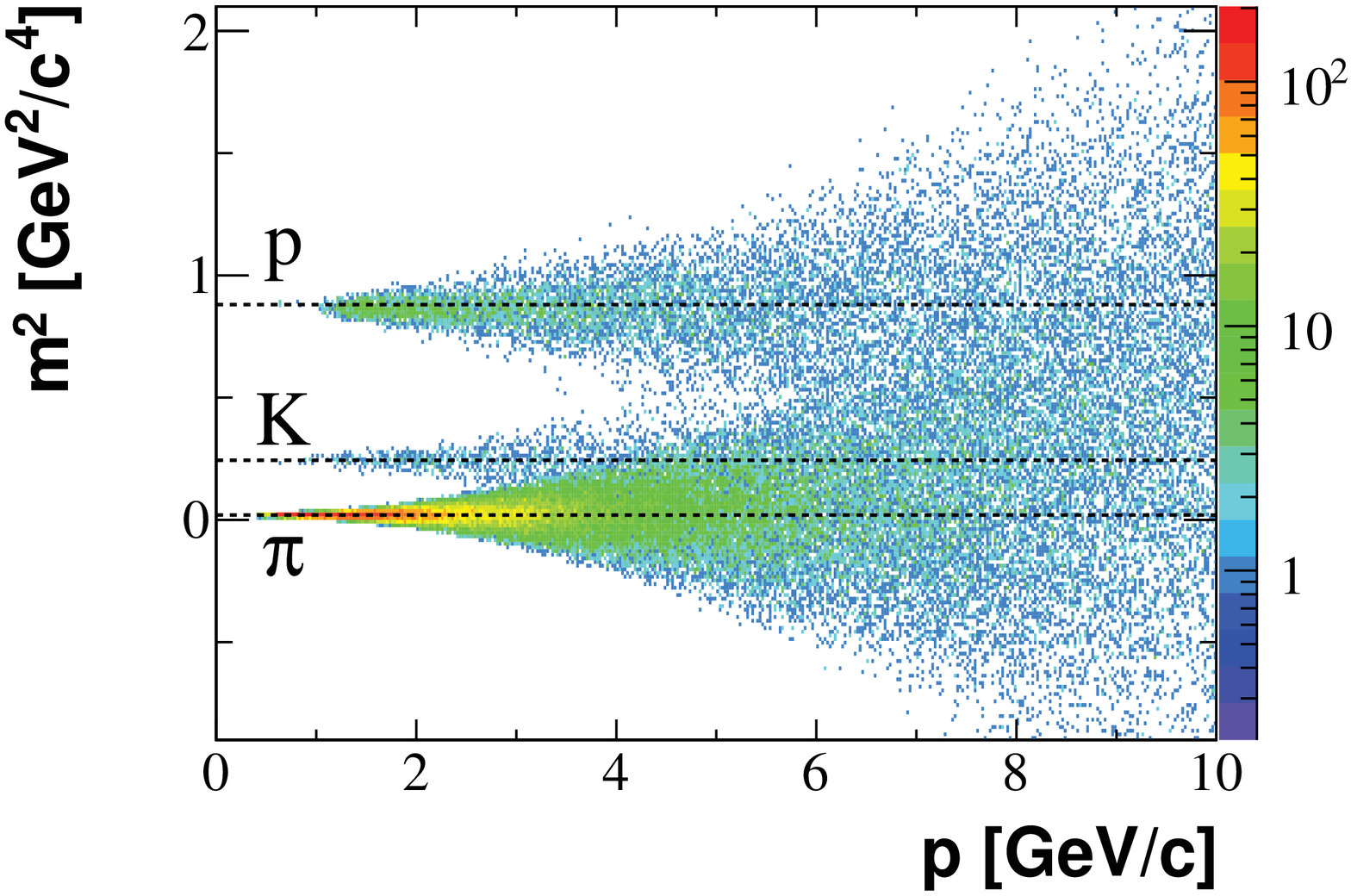,width=0.325\textwidth}
\epsfig{file=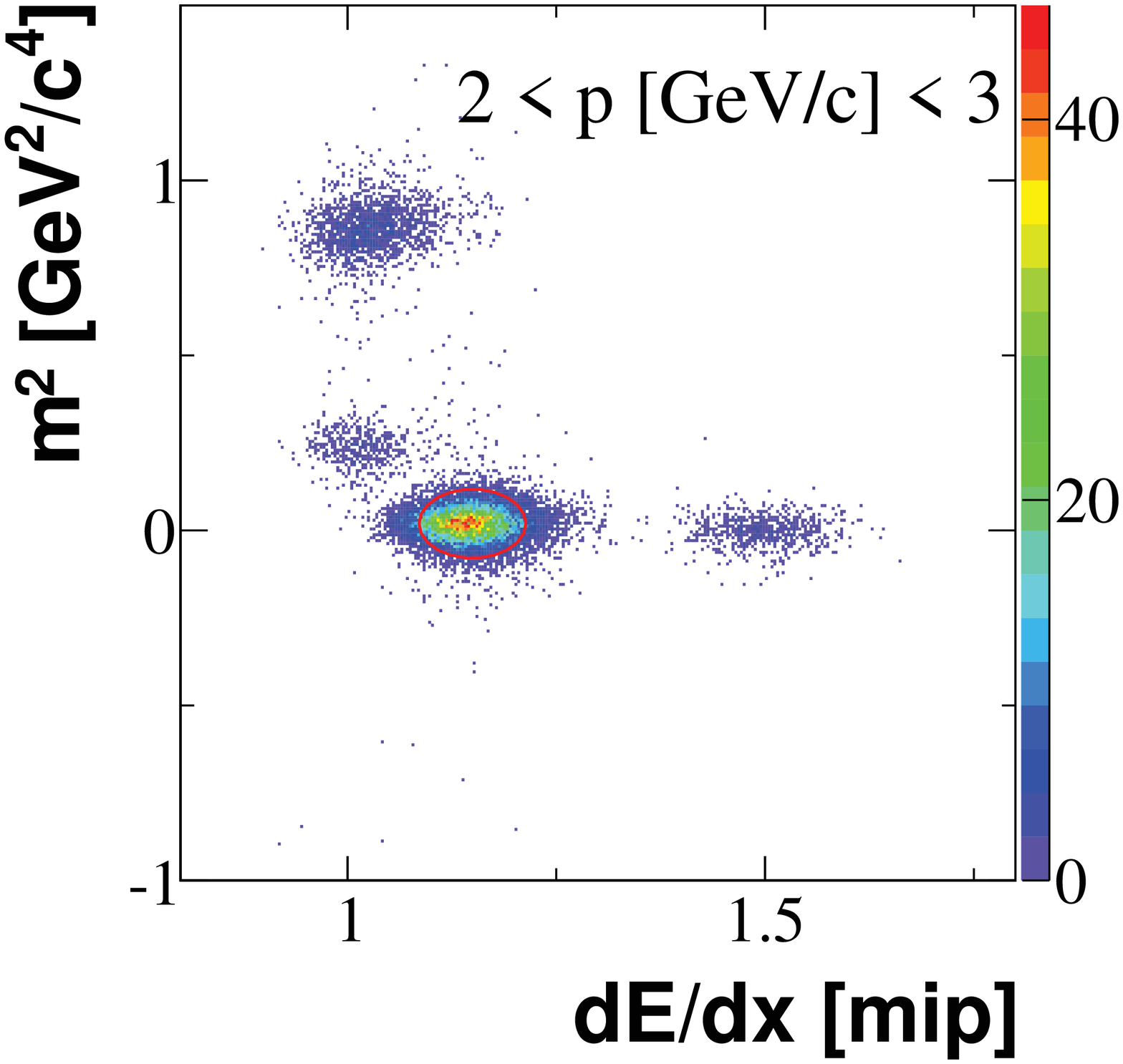,height=3.5cm,width=0.325\textwidth}
\caption{
\label{dedx_vs_m2}
Examples of PID capabilities of the NA61/SHINE spectrometer
for positively charged particles.
(Left)
Specific energy loss in the TPCs as a function of momentum. 
Curves show parameterizations of the mean $dE/dx$ calculated 
for different particle species.
(Middle)
Mass squared, derived from the ToF-F measurement and the fitted 
path length and momentum, versus momentum.
The lines show the expected mass squared values for different particles.
(Right)
Example of two-dimensional $m^2$--$dE/dx$ plot for
the momentum range 2-3~GeV/$c$.
Four clear accumulations corresponding to positrons, pions, kaons
and protons are observed.
$2\sigma$ contour around fitted pion peak is shown.
The combination of both $m^2$ and $dE/dx$
measurements provides close to 100\% purity in the pion
selection over the whole momentum range.
}
\end{figure}

The NA61 (SHINE) 
apparatus
is a large acceptance spectrometer
at the CERN SPS
for the study of the hadronic final states produced in interactions of
various beam particles ($\pi$, p, ions)
with a variety of fixed targets at the SPS energies.
The main components of the current detector
were constructed and used by the NA49 experiment~\cite{na49-nim}.
The main tracking devices are four large volume
TPCs.
Two of them, the vertex TPCs (VTPC-1 and VTPC-2), are
located in the
magnetic field of two super-conducting dipole magnets (maximum bending
power of 9~Tm)
and two others (MTPC-L and MTPC-R) are positioned
downstream of the magnets symmetrically with respect to the beam line.
One additional small TPC, the so-called gap TPC (GTPC), is installed on
the beam axis between the vertex TPCs.
The setup is supplemented by time of flight detector arrays
two of which (ToF-L/R) were inherited from NA49 and can provide
a time measurement resolution of $\sigma_{tof} \le 90$~ps.
A new forward time of flight detector (ToF-F) was
constructed in order to extend the acceptance of the NA61 (SHINE) setup
for pion and kaon identification
as required for the T2K measurements.
The particle identification in NA61 is based on the differential
energy loss dE/dx measured in the TPCs combined with the mass squared 
measurements based on the ToF information, see Fig.~\ref{dedx_vs_m2}.

Two carbon (isotropic graphite)
targets were used for T2K-related measurements:
1) a 2~cm-long target (about 4\% 
$\lambda_{\mathrm{I}}$) with density $\rho = 1.84$~g/cm$^3$,
the so-called thin target;
2) a 90 cm long cylinder of 2.6~cm diameter
(about 1.9 $\lambda_{\mathrm{I}}$),
the so-called T2K replica target with density $\rho = 1.83$~g/cm$^3$.

\subsection{First results}

The NA61 (SHINE) experiment was approved at CERN in June 2007.
The first pilot run was performed during October 2007.
In total 
about 670~k events with the thin target, 230~k events
with the T2K replica target and 80~k events
without target (empty target events) were registered during the 2007 run.
Using these data interaction cross sections and charged pion spectra
in p-C interactions at 31~GeV/c were measured~\cite{NA61_first_results}.
Such measurements are required to improve predictions of the
neutrino flux for the T2K long baseline neutrino oscillation 
experiment in Japan~\cite{t2k}.

For normalization and cross section measurements we adopted 
the same procedure as the one developed by 
the NA49 Collaboration~\cite{norm_NA49}. 
The measured inelastic and production cross sections 
in p-C interactions at 31~GeV/c are
257.2 $\pm$ 1.9 $\pm$ 8.9~mb and 229.3 $\pm$ 1.9 $\pm$ 9.0~mb, respectively.

\begin{figure}[tb]
\epsfig{file=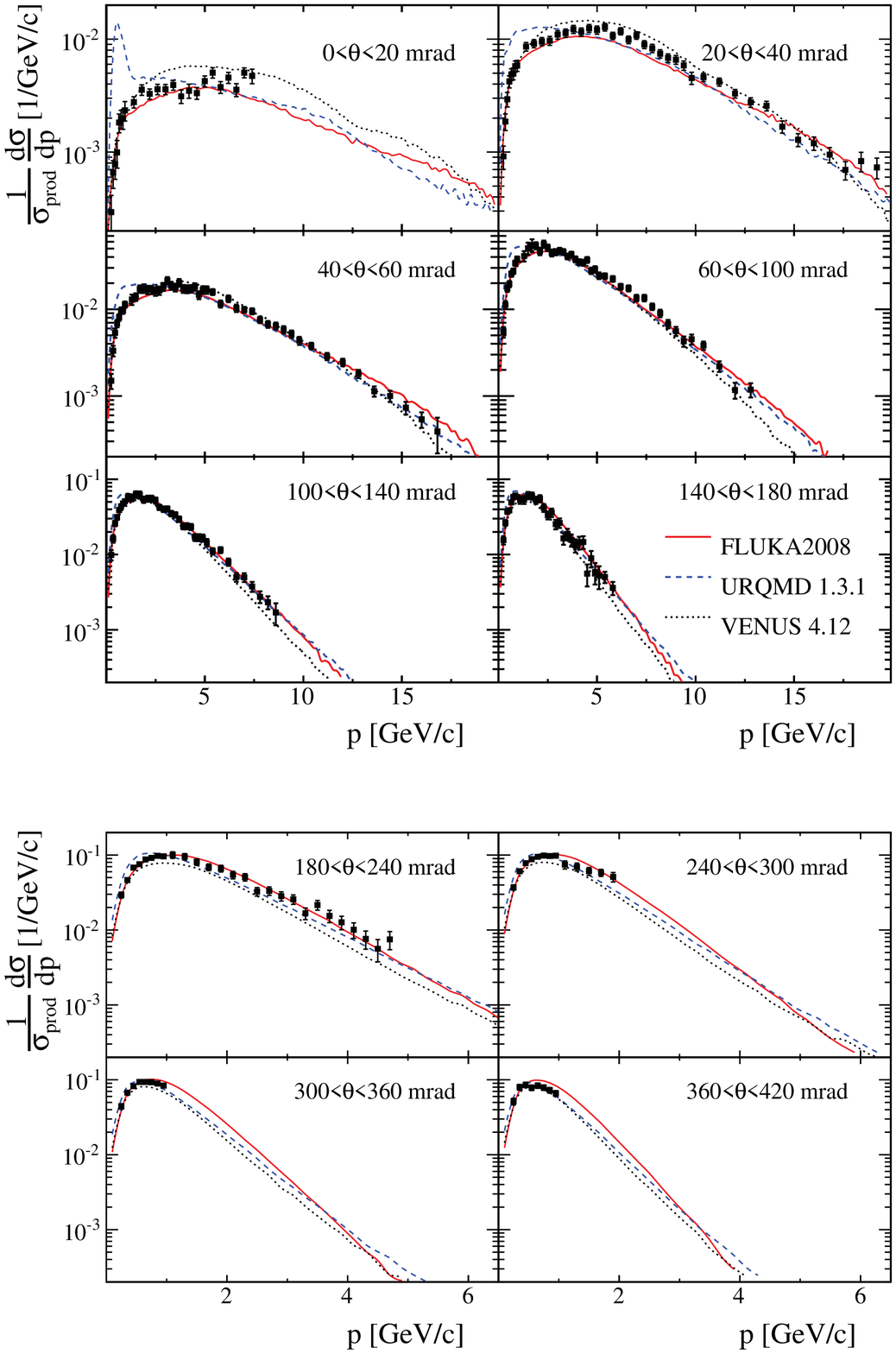,width=0.95\textwidth,height=11cm}
\caption{
\label{pion_plus_all_mult}
  Laboratory momentum distributions of $\pi^{+}$ mesons produced
  in production p-C interactions at 31~GeV/c
  in different intervals of polar angle ($\theta$).
  The spectra are normalized to the mean $\pi^{+}$ multiplicity in
  all production p-C interactions.
  Error bars indicate statistical and systematic uncertainties
  added in quadrature.
  The overall uncertainty~($2.3\%$) due to the normalization procedure 
  is not shown.
  Predictions of hadron production models,
  FLUKA2008 (solid line),
  URQMD1.3.1 (dashed line) and VENUS4.12 (dotted line) are also indicated.
}
\end{figure}

Crucial for this analysis is the identification of the produced charged pions.
Depending on the momentum interval, different approaches have been
adopted, which lead also to different track selection criteria.
The task is facilitated for the negatively charged pions, by the
observation that more than 90\% of primary negatively charged particles
produced in p-C interactions at this energy are $\pi^-$, and thus the analysis
of $\pi^-$ spectra can also be carried out without additional particle
identification.

Three analysis methods were developed to obtain charged pion spectra.
These are:
1) analysis of $\pi^-$ mesons via measurements of
negatively charged particles ({\it $h^-$ analysis});
2) analysis of $\pi^+$ and $\pi^-$ mesons
identified via $dE/dx$ measurements
in the TPCs ({\it $dE/dx$ analysis at low momentum}) and
3) analysis of $\pi^+$ and $\pi^-$ mesons
identified via time-of-flight and $dE/dx$
measurements in the ToF-F and TPCs, respectively
({\it $tof-dE/dx$ analysis}).
Each analysis yields fully corrected pion spectra with
independently calculated statistical and systematic errors.
The spectra  were compared
in overlapping phase-space domains to check their consistency.
Complementary domains were combined to reach maximum acceptance.

The agreement between spectra obtained by different methods is, 
in general, better than 10\%.
In order to obtain the final spectra consisting of
statistically uncorrelated points the measurement with the
smallest total error was selected.

Inclusive production cross sections for negatively and positively charged pions
are presented as a function of laboratory momentum in 10 intervals
of the laboratory polar angle covering the range 
from 0 up to 420~mrad~\cite{NA61_first_results}.
The final spectra for $\pi^+$ are plotted in Fig.~\ref{pion_plus_all_mult}.
For the purpose of a comparison of the data with model
predictions the spectra
were normalized to the
mean $\pi^+$ 
multiplicity in all
production interactions.
This avoids uncertainties due
to the different treatment of quasi-elastic interactions
in models as well as problems due to the absence of predictions for
inclusive cross sections.

As a first application of these measurements, it
is interesting to compare the $\pi^+$ spectra in p-C
interactions at 31~GeV/c to the predictions of event generators of
hadronic interactions.  
Models that have been frequently used for the interpretation of cosmic
ray data, i.e. VENUS4.12~\cite{Venus}, 
FLUKA2008~\cite{Fluka} and URQMD1.3.1~\cite{Urqmd}
were selected. They are part
of the CORSIKA~\cite{Corsika} framework for the simulation of air
showers and are typically used to generate hadron-air interactions at
energies below 80~GeV.  
In order to assure that all relevant settings of the
generators are identical to the ones used in air shower simulations,
p-C interactions at 31~GeV/c were simulated within CORSIKA in the
so-called {\itshape interaction test} mode.

The results are presented in Fig.~\ref{pion_plus_all_mult}.
The URQMD1.3.1 model qualitatively disagrees with the data
only at low momenta ($p <$~3~GeV/c) and polar angles below
about
140~mrad.
The VENUS4.12 and FLUKA2008 models follow the data trend in
all measured polar angle intervals.

The data presented here and in Ref.~\cite{NA61_first_results} 
already provide important information used
for improved calculation of the T2K neutrino flux. Meanwhile,
a much larger data set with both the thin (4\%~$\lambda_{\mathrm{I}}$)
and the T2K replica carbon targets
was recorded in 2009 (about 6~M triggers with the thin target and 2~M triggers 
with the replica target) and 2010 (about 10~M triggers with the replica 
target) and is presently being analyzed. This will lead
to results of higher precision for pions and extend the measurements to
other hadron species such as charged kaons, protons, $K^0_S$ and $\Lambda$. The
new data will allow a further significant reduction of the uncertainties
in the prediction of the neutrino flux in the T2K experiment.

\vspace*{-0.275cm}
\section{Conclusions}\label{sec:conclusions}

The HARP experiment has already made important contributions to the
cross-section measurements relevant for neutrino experiments. 

The MIPP experiment has nearly finalized measurements of hadron production
from the NuMI target used in the MINOS experiment.  

First measurements of charged pion production in p-C
interactions at 31~GeV/c
released recently by NA61 (SHINE) are of significant
importance for a precise prediction of the J-PARC neutrino beam
used for the first stage of the T2K experiment.
 
All three experiments provide also a large amount of input 
for validation and tuning of
hadron production models in Monte-Carlo generators.

Clearly, hardon production studies is a must for precision neutrino experiments.


\vspace*{-0.3cm}
\section{Acknowledgements}
It is a pleasure to thank the organizers of the ``Neutrino Telescopes'' 
in Venice for the 
invitation to participate in this very interesting conference 
and for the possibility to present 
the results described here. The author is grateful to colleagues from
the HARP and NA61/SHINE Collaborations.


\end{document}